# Graphene for Controlled and Accelerated Osteogenic Differentiation of Human Mesenchymal Stem Cells

*Tapas R. Nayak[1,†], Henrik Andersen[2,†], Venkata S. Makam[1], Clement Khaw[3], Sukang Bae[4], Xiangfan Xu[2], Pui-Lai R. Ee[1], Jong-Hyun Ahn[4,5], Byung Hee Hong[4,6], Giorgia Pastorin[1,7,8,*], Barbaros Özyilmaz[2,7,8,*]*

[1]Department of Pharmacy, National University of Singapore, 3 Science Drive 2, Singapore 117543; [2]Department of Physics, National University of Singapore, 2 Science Drive 3, Singapore 117542; [3]Nikon Imaging Centre, No. 11 Biopolis Way, Helios Building, Singapore 138677; [4]SKKU Advanced Institute of Nanotechnology (SAINT) and Center for Human Interface Nano Technology (HINT), Sungkyunkwan University, Suwon 440-746, Korea; [5]School of Advanced Materials Science and Engineering, Sungkyunkwan University, Suwon 440-746, Korea; [6]Department of Chemistry, Sungkyunkwan University, Suwon 440-746, Korea; [7]NUS Graduate School for Integrative Sciences and Engineering, Centre For Life Sciences (CeLS), 28 Medical Drive, #05-01, Singapore 117456; [8]NanoCore, Engineering Block A, EA, Level 4, Room No. 27, Faculty of Engineering, National University of Singapore, Singapore 117576.

†These authors contributed equally to this work.

*To whom correspondence should be addressed. Barbaros Özyilmaz: barbaros@nus.edu.sg, Giorgia Pastorin: phapg@nus.edu.sg.

**Abstract**

Modern tissue engineering strategies combine living cells and scaffold materials to develop biological substitutes that can restore tissue functions. Both natural and synthetic materials have been fabricated for transplantation of stem cells and their specific differentiation into muscles, bones and cartilages. One of the key objectives for bone regeneration therapy to be successful is to direct stem cells' proliferation and to accelerate their differentiation in a controlled manner through the use of growth factors and osteogenic inducers. Here we show that graphene provides a promising biocompatible scaffold that does not hamper the proliferation of human mesenchymal stem cells (hMSCs) and accelerates their specific differentiation into bone cells. The differentiation rate is comparable to the one achieved with common growth factors, demonstrating graphene's potential for stem cell research.

**Keywords:** Graphene; mesenchymal stem cells; cell differentiation; bone; osteogenesis.

Human mesenchymal stem cells (hMSCs) are critical for numerous groundbreaking therapies in the field of regenerative medicine. A myriad of environmental factors including their interaction with soluble growth factors, extracellular matrices and neighboring cells are crucial for their survival, proliferation and differentiation into specific lineages.[1-3] One of the main goals of tissue engineering is to control these factors by creating physical and chemical microenvironments designed to guide the ultimate fate of stem cells. Materials with different elasticity, rigidity and texture have been extensively investigated for this purpose. Stem cell scaffolds, which can be both 2D and 3D in nature, have been fabricated to mimic the intrinsic characteristics of natural substrates such as muscle, bone and cartilage.[4-6] Recently, both the lithographic patterning of suitable surfaces such as polydimethylsiloxane (PDMS),[7] polymethyl methacrylate (PMMA),[8] self-assembled titanium dioxide ($TiO_2$)[9] rod arrays and functionalized carbon nanotubes[10] have been explored. While there have been tremendous advances in this field, many challenges still remain. In particular in the field of bone tissue engineering, almost all artificial materials require the multiple administration of growth factors to promote hMSC differentiation. In addition, many approaches face challenges when it comes to scalability and compatibility with implants. For example, an alloplastic (non-biologic) material under mechanical stress may not respond in a similar way as the surrounding host bone, resulting in structural failure of the implant or inflammatory changes in the original bone, as seen in stress shielding.[11] Also, bioactive implants still face limitations in terms of potential pathogenic infections, low availability and high costs. Graphene[12] may provide an elegant solution to some of these challenges. Being only one atom thick, it introduces the least amount of artificial material possible and has a large number of remarkable properties.[13] In the context of tissue engineering, its mechanical properties are likely to play a key role: graphene has the highest Young's modulus (0.5 – 1 TPa) among any known material, yet it is not brittle.[14,15] Graphene can be transferred onto any flat or irregular shaped surface and graphene–coated, flexible, supporting substrates can be easily bent into any shape required.[16]

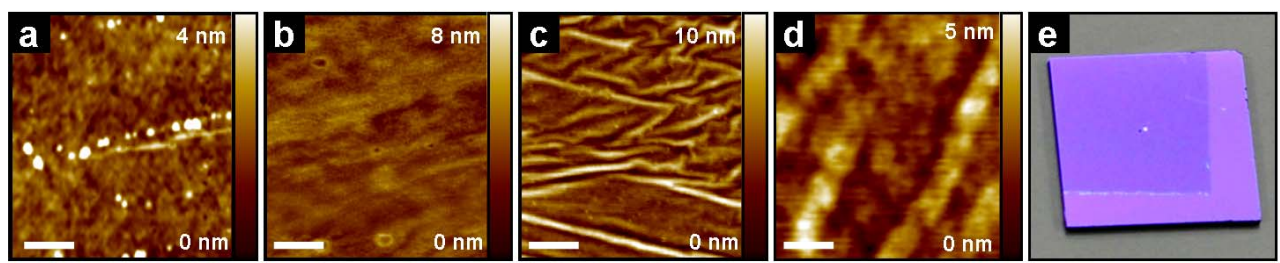

**Figure 1**. Graphene onto different substrates. AFM of graphene on (**a**) Si/SiO$_2$ (**b**) Glass slide (**c**) PET-film and (**d**) PDMS. Scale bars are 200nm (**e**) Optical image of 1cm x 1cm, partially graphene coated Si/SiO$_2$ chip, showing the graphene boundary.

Stem cell research with graphene has become feasible only with the recent availability of cheap, high-quality, continuous graphene sheets on a large scale.[17] Here we show that graphene provides a new type of biocompatible scaffold for stem cells. Remarkably, graphene accelerates cell differentiation even in the absence of commonly used additional growth factors such as BMP-2. Taking into consideration both the intrinsic mechanical properties of graphene and the striking results of this study, we envisage a functional role of this new material as a versatile platform for future biomedical applications in general and stem cell therapies in particular.

Large-scale graphene used in this study was synthesized by the chemical vapor deposition method on copper foils. After growth, copper was etched and the same batch of graphene was transferred to four distinct substrates used in this study according to methods discussed elsewhere.[18] We studied the influence of graphene on stem cell growth by investigating four substrates with widely varying stiffness and surface roughness: (1) polydimethylsiloxane (PDMS), (2) polyethylene terephthalate (PET), (3) glass slide and (4) silicon wafer with 300nm $SiO_2$ ($Si/SiO_2$) (Table S1 in Supporting Information). Plain cover slips without graphene were used as a control or reference for normalization. Atomic Force Microscopy (AFM) was used to analyze the surface roughness of the various substrates with and without graphene coating. Transferred to PET and PDMS, the graphene sheet exhibits nano-ripples with high density (Fig. 1c,d) compared to graphene on $Si/SiO_2$ and glass slide (Fig. 1a,b). Despite being only one atom thick, on $Si/SiO_2$ substrates with well-defined oxide thickness, graphene can be easily seen with a simple conventional optical microscope (Fig. 1e). Therefore, detailed studies such as the evolution of cell differentiation with time were done mainly on graphene coated $Si/SiO_2$ substrates. Two distinct sets of experiments were performed. First cell viability was studied with cells cultured in normal stem cell medium. Next, stem cell differentiation was examined in cells cultured on conventional osteogenic media.

## Results and Discussion

**Cell viability and morphology**

We first discuss cell morphology and viability by image analysis on all four substrates with and without graphene coverage when cells were cultured in normal stem cell media. From Fig. 2a we see that, independent of the substrate, there is no significant difference (p>0.05) in cell viability between graphene-coated and uncoated substrates. We also performed MTT assays (Fig. S2 in SI) to confirm the cell viability data. Again, regardless of the substrate, there was no difference (p>0.05) between uncoated and graphene-coated substrates, demonstrating that cell growth was indeed not affected by the presence of graphene. Note that cell viability is lower on PET and PDMS independent of the presence of graphene.

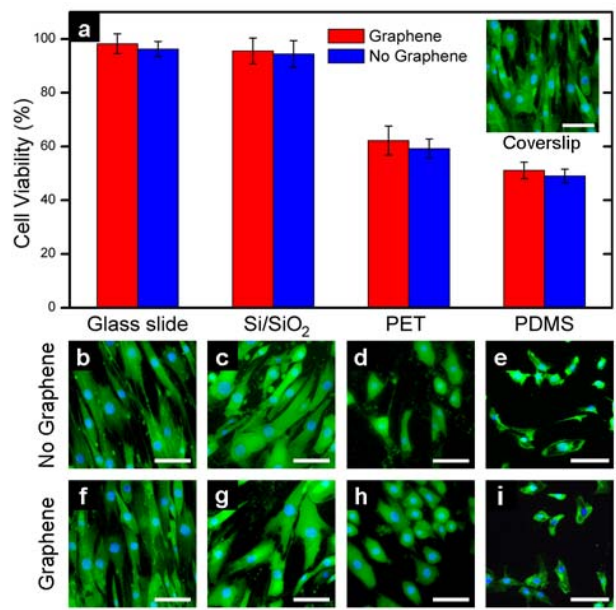

**Figure 2.** Cell viability and morphology of hMSCs grown on different substrates. Cells are stained with DAPI (blue) and Calcein AM (green). **(a)** Graph showing cell viability in percentage normalized to cover slips used as reference. **(Inset)** Morphology of hMSCs grown on standard cover slips. **(b-e)** hMSCs grown on glass slide, Si/SiO$_2$, PET and PDMS without graphene. **(f-i)** hMSCs grown on graphene coated glass slide, Si/SiO$_2$, PET and PDMS. Scale bars are 100μm.

A similar conclusion can be reached by just comparing cell morphology with and without graphene. In general, the presence of graphene (Fig. 2f-i) did not influence the shape of the cells in comparison to uncoated substrates (Fig. 2b-e). Mesenchymal stem cells maintained their spindle-shape across glass slides and Si/SiO$_2$ after 15 days of incubation (Fig. 2b,c and 2f,g). Here stem cells presented the usual elongated structure with noticeable filopodia extensions and cellular propagation fronts. In the case of PET and PDMS, cells showed rounded or irregular morphology, most probably due to poor adhesion to the substrate (Fig. 2d,e and 2h,i). This suggests that graphene does not hamper the normal growth of stem cells and that the incorporation of this material in implants or injured tissues would not affect the physiological conditions of the microenvironment. In fact, Raman measurements and visual inspection of the samples after cell incubation and subsequent removal clearly showed that the graphene sheets remained largely intact (Fig. S5 in SI).

**hMSCs differentiation into osteogenic lineages**

Next, specific markers were used to determine the conversion of hMSCs into specific cell types when cultured in osteogenic media. Note that conventional osteogenic medium does contain dexamethasone, which can lead to osteogenic differentiation by itself. However, it is usually administered in combination with other agents and growth factors such as BMP-2 to achieve differentiation through a synergistic effect. In none of substrates studied here, the osteogenic medium alone was sufficient to lead to osteogenic differentiation over the whole duration of the experiment (15 days). In the absence of graphene, stem cells on cover slips, on glass slides (not shown) and on Si/SiO$_2$ (Fig. 3a,b) did not differentiate: this was demonstrated by immunofluorescent staining of two typical protein markers, namely CD-44 for hMSCs and osteocalcin (OCN) for osteoblasts. These three substrates showed a CD-44-positive staining and the absence of OCN. However, once these stiff substrates were coated with graphene, hMSCs lost their ability to bind the fluorescent antibody specific for CD-44 expression, suggesting they underwent a different fate (Fig. 3c). In fact, hMSCs immunostained for OCN (Fig. 3d), indicating osteogenic differentiation. On uncoated PDMS, hMSCs did not stain CD-44 but they showed a weak expression of MAP2 (typical neuronal marker, Fig. 3e). On the other hand, in the case of uncoated PET, desmin (D33, a muscle cell marker) staining but not CD-44 was observed (Fig. 3i). However, once coated with graphene, hMSCs growing also on these softer substrates bound specifically to OCN (Figs. 3h and 3l) only, demonstrating that graphene is the driving force of bone cell formation, regardless of the underlying substrate.

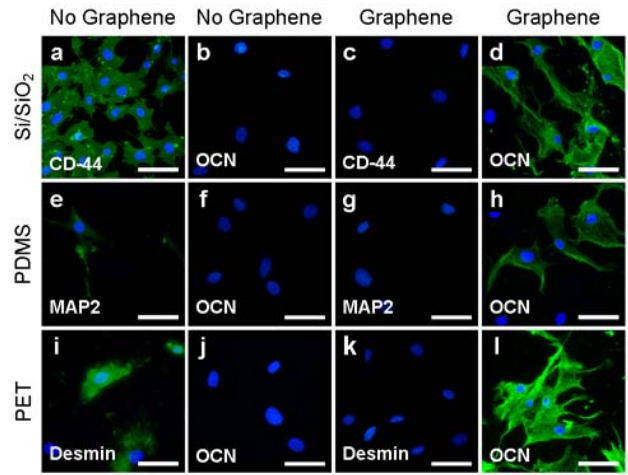

**Figure 3.** Immunostaining of cells growing on Si/SiO$_2$, PDMS and PET without BMP-2 growth factor. Cells are stained with DAPI (blue) and either CD-44, MAP2, Desmin or Osteocalcin (OCN) as indicated (green). **(a-d)** Cells growing on Si/SiO$_2$, without graphene showing presence of CD-44, and with graphene showing presence of OCN. **(e-h)** Cells growing on PDMS without graphene showing some MAP2 immunostaining, and with graphene showing staining of OCN. **(i-l)** Cells growing on PET without graphene showing some staining of desmin, and with graphene showing OCN immunostaining. Scale bars are 100 μm.

This is most clearly seen in Fig. 4b, showing the immunofluorescent staining of cells on a Si/SiO$_2$ wafer, which are cultured in osteogenic medium but only partially covered by graphene. Despite the stiffness of the substrate, specific immunostaining for OCN was only observed in the area covered by graphene. The boundary separating the graphene coated region from the uncoated region is clearly visible even from the immunofluorescent image. These qualitative observations have been confirmed by quantitative alizarin red staining (Fig. 4c,d), which indicates the presence or absence of calcium deposits due to bone nodule formation. The results for all substrates are summarized in Fig. 4c, where we compared the extent of calcium deposition on each substrate, with and without graphene coating, in the absence of the typical growth factor BMP-2. A strong increase in calcium deposit with graphene coating is observed for all substrates. While the effect is more pronounced with the stiffer substrates, surprisingly graphene had a similar effect also on the softer substrates PET and PDMS. It should be noted that in the absence of growth factors both PDMS and PET are known to be less favorable towards osteoblasts.[7] Yet the presence of graphene induced a drastic change of their natural behavior similar to what has been observed with apatite coating on such polymers.[20-22] We would like to emphasize that also here the osteogenic medium alone was not sufficient to induce differentiation: within the 15 day time frame of the experiment, the control represented by cover slips in osteogenic medium *without* graphene, *i.e.* hMSC cultured on ordinary tissue culture plate, did not show any calcium deposition. The evaluation of a separate set of samples by fluorescent-activated cell sorting (FACS) further supported our findings that graphene does accelerate the osteogenic differentiation of stem cells (Fig. S3 in SI).

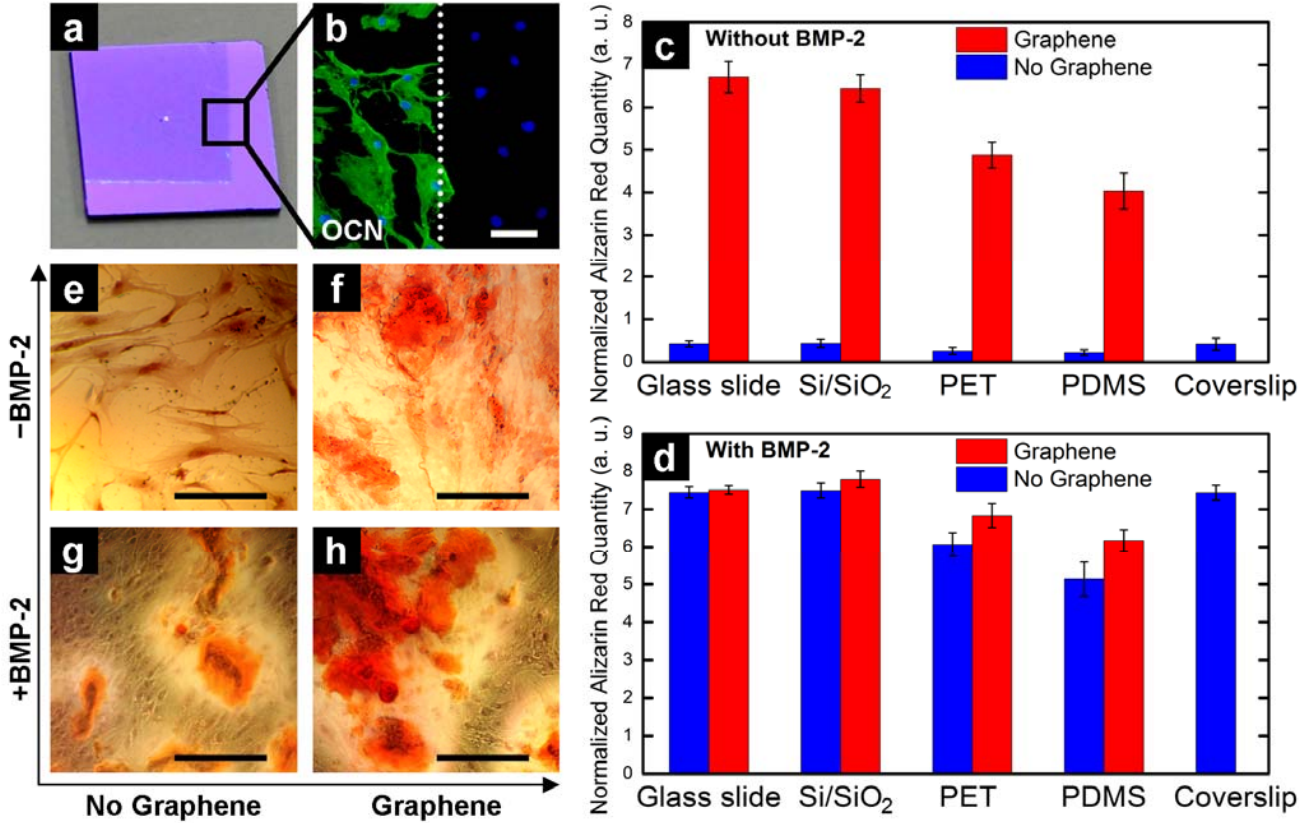

**Figure 4.** Graphene accelerates osteogenic differentiation. (**a**) Optical image of 1cm x 1cm, partially graphene coated Si/SiO$_2$ chip, showing the graphene boundary. (**b**) Osteocalcin (OCN) marker showing bone cell formation on the same chip only on the graphene coated area; the graphene boundary is also clearly visible here. (**c-d**) Alizarin Red quantification deriving from hMSCs grown for 15 days on substrates with/without graphene. (**c**) Cells grown in the absence of BMP-2. Control with cover slips is shown as a reference. (**d**) Cells grown in the presence of BMP-2. Conventional plain cover slips were used as a positive control. (**e-h**) PET substrate stained with alizarin red showing calcium deposits due to osteogenesis.

(**e**) PET without BMP-2 and without graphene; (**f**) PET without BMP-2 and with graphene; (**g**) PET with BMP-2 and without graphene; (**h**) PET with both BMP-2 and graphene. Scale bars are 100μm.

The impact of graphene on softer substrates such as PET became even more evident in a parallel study, in which we directly compared graphene's influence to that of BMP-2 (Fig. 4e-h) after 15 days of incubation. In the absence of both graphene and BMP-2, no bone nodule formation was observed as indicated by negative alizarin red staining (Fig. 4e). As expected we see positive staining with identical samples after the addition of BMP-2 (Fig. 4g). On the other hand graphene-coated PET showed a positive staining even without BMP-2 (Fig. 4f). We also performed experiments where we combined both graphene coating and BMP-2 treatment (Fig. 4h). In the case of PET and PDMS we observed a significant enhancement of calcium deposits compared to the above-mentioned samples, which were either only coated with graphene or only treated with BMP-2. This enhancement was specific to soft substrates, and much less evident on the stiffer glass slides and Si/SiO$_2$. Finally, quantitative alizarin red staining was used to study the role of graphene in the presence of BMP-2 (Fig. 4d) for all four substrates. On the stiffer substrates (*i.e.* glass slide and Si/SiO$_2$) the additional presence of graphene did not further enhance the production of calcium deposits ($p>0.05$). In fact calcium deposits had almost saturated purely from graphene, deeming the biochemical growth factor unnecessary. On the other hand a clear, statistically significant increase ($p<0.005$) is seen on the softer materials PET and PDMS. This again suggests that graphene itself has a remarkable role in the differentiation of hMSCs towards the osteogenic lineage.

**Time-dependent study of differentiation**

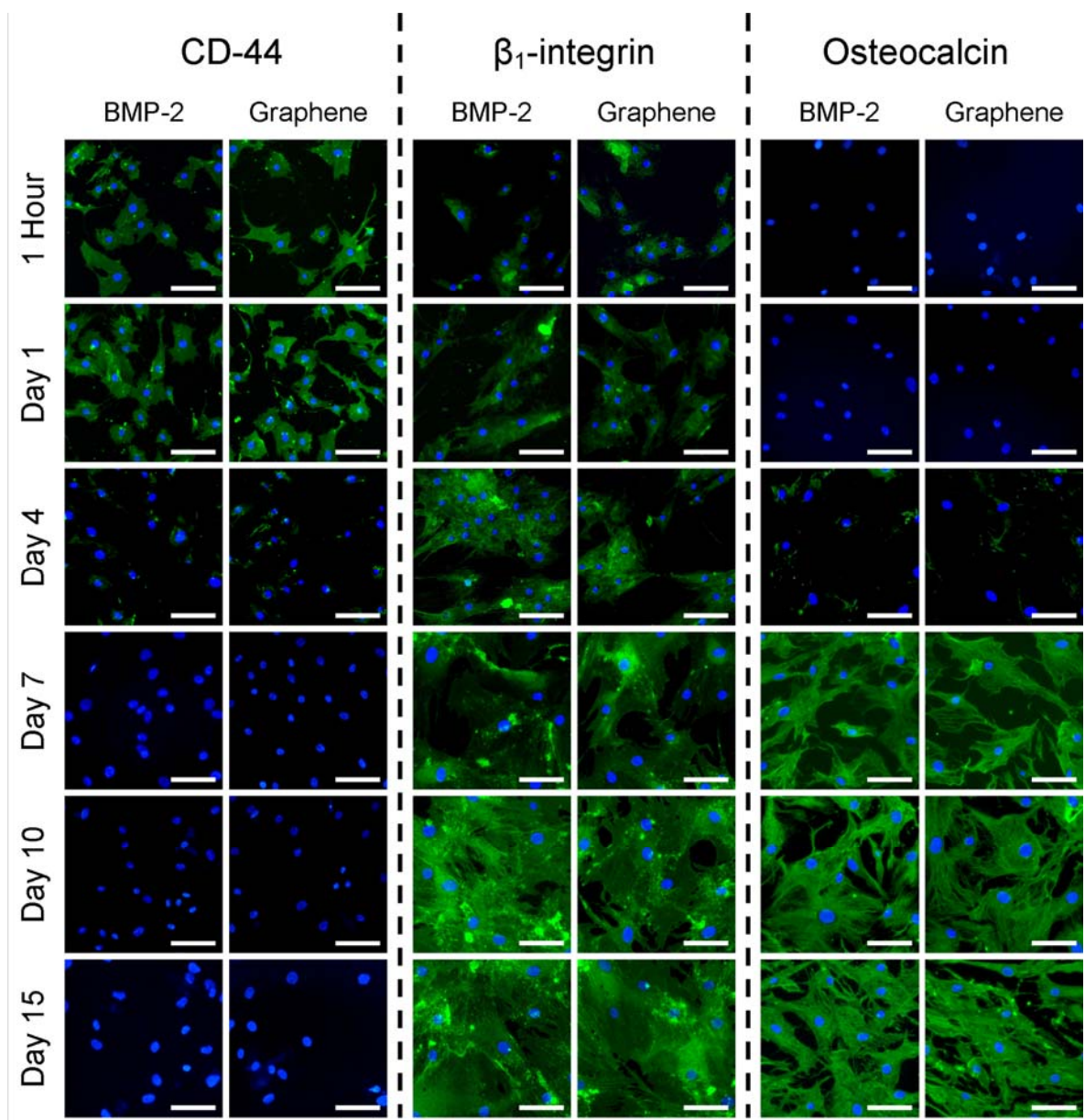

**Figure 5.** Immunostaining of cells growing on Si/SiO$_2$ substrates either treated with BMP-2 or coated with graphene. Experiments were performed from 1 hour to 15 days. (**left**) CD-44, marker for stem cells, decreased over time and completely disappeared by Day 7. (**center**) β1-integrin, marker for cell-substrate adhesion, increased over time, reaching its peak by Day 15. (**right**) OCN, marker for bone cells, became visible at Day 4 and very intense by DAY 7. Scale bars are 100μm.

An important parameter for practical applications is also the time a material takes to induce bone cell differentiation. To that purpose we studied how fast cells on graphene-coated Si/SiO$_2$ substrates differentiate over a time frame of 15 days in comparison to cells growing on uncoated Si/SiO$_2$, but treated with BMP-2. We studied these samples at specific time points of 1 hour and 4, 7, 10 and 15 days (Fig. 5). Interestingly, both BMP-2-treated and graphene coated substrates were able to induce cell differentiation at the same rate. More precisely, hMSCs on neither substrate showed any sign of osteoblast formation until Day 4. This is demonstrated by the intensity of fluorescence due to CD-44 marker, which is characteristic for stem cells and clearly visible already after one hour of incubation (DAY 1). Conversely, fluorescence due to CD-44 decreased remarkably by DAY 4 and completely disappeared by DAY 7. On the other hand, a progressive enhancement of fluorescence was observed due to OCN (indication of terminal osteogenic differentiation) and $\beta_1$-integrin, a protein indicating cell-substrate interaction. The results confirmed a successful differentiation into bone cells with a strong adhesion to the substrates by DAY 7 for both types of samples. $Si/SiO_2$ substrates treated with a) only BMP-2 and b) only graphene were able to accelerate cell differentiation at the same rate over a period of 15 days of incubation (Fig. 5). Equally important, in contrast to graphene, BMP-2 needed to be administered every three days during the course of the experiment due to the very short half-lives of BMP-2[23,24] again showing graphene as a worthy replacement of biochemical growth factors.

**Control Experiments**

To confirm that graphene is critical for the observed stem cell differentiation, we performed control experiments with both amorphous carbon thin films and highly oriented pyrolytic graphite (HOPG) samples. Following identical experimental protocols, we observed that while both types of samples did support cell proliferation, none of them led to cell differentiation (Fig. 6d,f and Fig. S4 in SI). Here we discuss the HOPG results in more detail. The results on amorphous carbon thin films are summarized in SI.

The AFM images of graphene and HOPG (Fig. 6a,b) clearly show the difference in their topography. Cells were cultured on graphene or HOPG in osteogenic medium. After 4 days the fluorescence deriving from the antibody specific for CD-44 expression was significantly lower for cells grown on graphene (Fig 6c) than for cells on HOPG (Fig. 6d). At the same time, specific immunostaining for OCN is already detectable with cells grown on graphene (Fig 6e), while only the DAPI stained nuclei are visible for cells on HOPG (Fig. 6f).

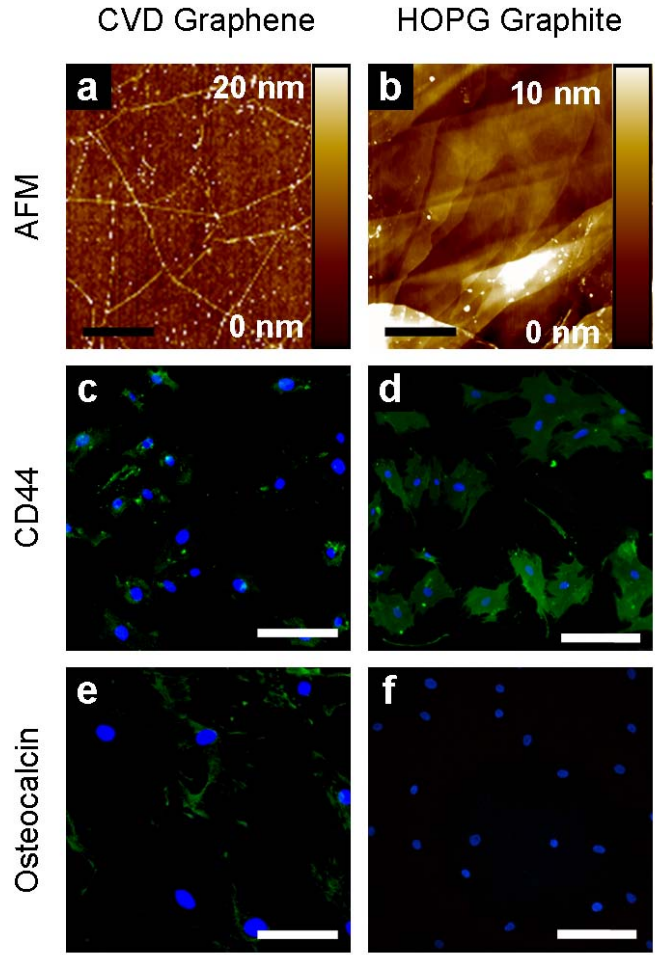

**Fig. 6**: Comparison of osteogenic differentiation of hMSCs after 4 days on graphene-coated Si/SiO$_2$ and HOPG on cover slips. AFM images of (**a**) graphene and (**b**) HOPG. Scale bars are 5μm. (**c-d**) Immunostating of CD-44 (**green**) on graphene and HOPG, respectively. (**e-f**) Immunostaining of OCN (**green**) on graphene and HOPG, respectively. Nucleus is stained with DAPI (**blue**). Scale bars are 100μm.

It is rather remarkable that a single continuous sheet of carbon atoms can strongly accelerate bone cell differentiation. The observed effect is almost certainly due to a complex interplay of mechanical, chemical and electrical properties of graphene and the interactions between graphene and cells, as well as graphene and supporting substrates. This makes it difficult to identify the microscopic origin of the effect. However, a comparison with related CNT-based experiments does offer some clues (see SI). In addition, control experiments with amorphous carbon thin films and HOPG samples do provide some insights. In particular, the disparities between the results obtained with graphene and HOPG point towards mechanical properties and surface morphology as the decisive factors. While locally (~100nm) the two systems have comparable surface morphology, on a larger scale they look very different. Graphene consists of many ripples and wrinkles on the micron scale (Fig. 6a). Such localized out-of plane deformations are completely absent in HOPG graphite, the surface of which consists instead of a large number of micron size terraces (Fig. 6b).

The correlation of cell morphology and substrate morphology has been investigated by several groups. Dulgar-Tulloch *et al.*[25] cultured hMSCs on ceramics with varying grain sizes between 24nm to 1500nm showing, that the 200nm grain size was most favorable for hMSC proliferation independent of the surface chemistry, the surface roughness, and the crystal phase. Oh *et al.*[9] studied hMSCs on various diameters of $TiO_2$ nanotubes and observed a decrease in adhesion, but an increase in osteogenic differentiation with increasing nanotube diameter. This is in good agreement with cell morphology induced differentiation observed by McBeath *et al.*[26] Even nanoscale patterning of poly(methyl methacrylate) (PMMA) has been reported to induce bone cell differentiation. Surprisingly, cell differentiation mainly took place when the nanopit arrays were disordered.[8] While the microscopic origin of these effects are not yet fully understood, it is worth noting that the topography of CVD graphene with its many ripples and wrinkles does mimic the disordered nanopit array of Dalby *et al.*[8] (Fig. 6b). Such ripples and wrinkles are intrinsic to CVD graphene and originate from the difference in the thermal expansion coefficient of Cu and graphene.[18] Similarly to Dalby *et al.*,[8] the large scale disorder in CVD graphene could play a role in protein adsorption, cell adhesion, proliferation and differentiation. The ripples themselves also provide local curvature and hence, could further enhance the reactivity of such graphene sheets.

Cell differentiation depends strongly also on substrate stiffness[27] and strain, such that when applied (cyclically) it increases the expression of osteogenic markers for osteopontin (OPN) and BMP-2.[28,29] Therefore, graphene's exceptionally high Young's modulus[30] and its remarkable flexibility for out-of plane deformation could also contribute to stem cell differentiation. However, we expect a significant influence of graphene's Young's modulus only with substrates, which are much thinner and softer than used in our studies. Even in the case of the PDMS substrates, the presence of graphene will only marginally increase its Young's modulus. On the other hand, the ability of graphene to sustain lateral stress could play a more important role in the context of providing just the right amount of local cytoskeletal tension. Together with the observation that graphene allows for easy out-of-plane deformation, this may lead to the formation of strong anchor points of the cytoskeleton. Such tension may allow for the unfolding of the mechanically sensitive protein of interest and change conformation.[6] Such forces could potentially be even easier to realize on PET and PDMS substrates, since on these materials graphene is already pre-strained.[31] Cell induced out-of-plane deformations could further enhance the reactivity of nanorippled graphene sheets.

The fact that (HOPG) graphite is made out of weakly bound graphene planes may be equally important. In the presence of lateral forces such materials easily shear off and are therefore, commonly used in lubricants. In the specific context of cell adhesion and in view of the (lateral) contractual forces cells exert on the surface, this effect may hamper strong cell adhesion. Note that cells can mechanically "sense" lower lying layers down to several tens of micrometers. In the case of graphene, the cells sense the underlying (amorphous) substrate instead.

**Conclusions**

To summarize, the presence of graphene did not influence the shape and the growth of the cells in normal stem cell media, demonstrating biocompatibility and suggesting that the incorporation of this material in implants or injured tissues would not affect the physiological conditions of the microenvironment. In the presence of an osteogenic medium, graphene-coating helped by remarkably accelerating the differentiation of hMSCs at a rate comparable to differentiation under the influence of BMP-2. This represents a critical aspect to its successful use for stem cell-based regenerative medicine strategies. In contrast to other substrates, graphene is neither brittle nor does require further nanoscale patterning or functionalization. In addition it is scalable and provides a cost effective way to prepare scaffolds for biological tissues. Currently graphene is only available in form of sheets and we envision a promising role of graphene located between implants and the surrounding tissues. However, the conditions under which graphene is grown are being constantly improved. There is for example a strong effort in establishing graphene growth at much lower temperatures. Thus, growth on alternative biocompatible and biodegradable surfaces and 3D scaffolds, even without the need to resort to catalytic metal films, have recently demonstrated to be feasible.[32]

## Methods

**Substrate preparation**

Graphene was grown on copper foils by chemical vapor deposition at 1000°C in a mixture of hydrogen and methane as discussed elsewhere.[18] The graphene film was mechanically supported by a thin film of PMMA (Microchem) and the copper foil was etched in a weak solution of ammonium persulfate (Sigma). The graphene coated with PMMA was transferred to deionized water to remove residues and the transfer was completed by gently contacting graphene with the substrate and lifting it out of the water. To avoid any residues from the transfer process the samples were left in warm acetone for 12 hours followed by 3 hours in isopropanol. In a final step the $Si/SiO_2$ substrates were annealed in $Ar/H_2$ 90/10 wt% for 7 hours at 300°C to further reduce impurities in the graphene layer. However, note that $Si/SiO_2$ without the additional step of annealing showed the same cell viability and induced stem cell differentiation at the same rate (data not shown).

**Cell lines and markers**

Human mesenchymal stem cells (hMSCs) were purchased from ATCC and cultured in low-glucose Dulbecco's modified eagle medium (Sigma) supplemented with 10% FBS (Invitrogen), 1% penicillin/streptomycin (Gibco), 1% Non-essential amino acids (Sigma) and 1% sodium pyruvate (Sigma). hMSCs at passage 2 were used in this study. Osteogenic medium consisting of DMEM basal medium (Sigma) added with dexamethasone, L glutamine, ascorbic acida and Beta-glycerophosphate was prepared according to a known procedure.[33] FITC-Goat anti mouse antibody was purchased from Biolegend, San Diego, California (USA). Markers (osteocalcin (OCN), CD44, Desmin (D33), MAP-2, $β_1$-integrin) were purchased from Acris Antibodies GmbH (Germany).

**Cell Viability**

hMSCs (20,000 cells/well (24 well plate)) were seeded on uncoated (control) and graphene coated (test) chips and cultured in normal stem cell medium. Post confluence (2 weeks), cells growing on each chip were transferred to new well plate and washed 3 times with 2 ml of PBS. 1 ml of PBS was added to each well followed by 5μl of 1mM Calcein acetoxymethyl ester (Calcein AM) and incubated at room temperature for 15 minutes. After removing the unbound stains, the chips were inverted on to glass slides mounted with vectashield with DAPI (H 1200, Vector labs) and visualized under fluorescence microscope (Nikon AZ-100 multipurpose microscope). Pictures were taken at 40 different positions of the chips and processed by image J software to count the number of viable cells to the number of nucleus as determined by staining with DAPI. Cell viability was measured by comparing the cell numbers for each substrate with the cells counted on cover slips. In addition, (3-(4,5-Dimethylthiazol-2-yl)-2,5-diphenyltetrazolium bromide (MTT) assays were carried out, in which cytotoxicity evaluation was based on the activity of enzymes to reduce MTT to formazan dyes, giving a purple colour.[34] Experiments were carried out in triplicates, following the procedure reported in supporting document. The morphology of the hMSCs on different substrates were compared according to the image as seen in the form of calcein AM staining.

**Alizarin red staining and quantification**

hMSCs (20,000 cells/ well (24 well plate)) were seeded in to the control and the test well plate. After 24 hours, osteogenesis was induced by replacing the original medium with osteogenic medium, which was changed every 3 days up to confluence (2 weeks).

Alizarin red staining was performed using the protocol adapted from Chemicon Mesenchymal Stem cell Osteogeneis kit Cat. No. SCR028. Briefly, the medium was aspirated out from each well and cells were fixed with ice cold 70% ethanol for 1 hour at room temperature. Then the cells were rinsed twice with milliQ water followed by addition of 2 ml of alizarin red (Sigma) solution for each well and incubated for 30 minutes. Finally the unstained alizarin red was washed with milliQ water and the chips were visualized under microscope (Nikon eclipse TE2000-U, Japan). Cells with calcium deposits due to bone nodule formation were stained red. Alizarin red quantification was done using a previously reported procedure.[35]

**Immunofluorescence of hMSCs and time-dependent differentiation study**

hMSCs at 20,000 cells/well (24 well plate) were seeded, osteoinduced and incubated up to confluence (2 weeks) as reported above. The cells on all the chips were fixed by treating them with ice cold 50%/50% methanol/acetone. After 5 minutes, methanol/acetone was removed and the chips were left open inside the laminar hood to be air dried. After the chips were completely dried, the fixed cells were treated with 10% FBS (blocking agent) in PBS for 20 minutes. The blocking agent was aspirated out and 5 μl of different antibodies to cellular markers (CD-44 for hMSCs, OCN for osteoblasts, Desmin for muscle cells and MAP2 for neuronal cells) were added on to separate chips (previously marked). After 1 hour the cells on the chips were extensively washed in milliQ water for 5 minutes and then rinsed in PBS 1X for 5 minutes. After that, 100 μl of diluted (1/100) FITC-goat antimouse antibody were added on to each chip and incubated at room temperature. After 30 minutes the cells were washed with milliQ water for 5 minutes and then rinsed in PBS 1X for 5 minutes. The chips were inverted on to glass slides mounted with vectashield with DAPI (H 1200, Vector labs) and visualized under fluorescence microscope (Nikon AZ-100 multipurpose microscope).

For time-dependent differentiation experiment, osteogenic differentiation was further evaluated over a time frame of two weeks. Uncoated substrates were subjected to BMP-2 (75 ng/mL, added every 3 days) and compared to graphene coated substrates at 1 hour and at Day 1, 4, 7, 10 and 15 in terms of binding to CD-44 (which stains hMSCs), $β_1$-integrin (which indicates cell-substrate adhesion) and OCN (which indicates bone cells). The above mentioned procedure was followed for the immunofluorescence and imaging purposes.

B. Ö. acknowledges support by the Singapore National Research Foundation under NRF RF Award No. NRFRF2008- 07, and A*STAR SERC TSRP - Integrated Nano-Photo-Bio Interface Award and by NUS NanoCore. G. P. acknowledges the Singapore Bioimaging Consortium, Nikon Imaging Centre, for providing technical assistance and microscope facilities.

# Supporting Info

# Graphene for Controlled and Accelerated Osteogenic Differentiation of Human Mesenchymal Stem Cells

*Tapas R. Nayak[1,†], Henrik Andersen[2,†], Venkata S. Makam[1], Clement Khaw[3], Sukang Bae[4], Xiangfan Xu[2], Pui-Lai R. Ee[1], Jong-Hyun Ahn[4,5], Byung Hee Hong[4,6], Giorgia Pastorin[1,7,8,\*], Barbaros Özyilmaz[2,7,8,\*]*

# Substrates

Figure S1 shows AFM images of graphene on Si/SiO$_2$. The graphene surface topography consists of flat areas (<1nm RMS) separated by ~8 nm high ripples, which are caused by the difference in thermal expansion of the Cu foil (on which graphene is grown) and the graphene sheet itself (Li et al., *Science*, **2009**, *324*, 1312).

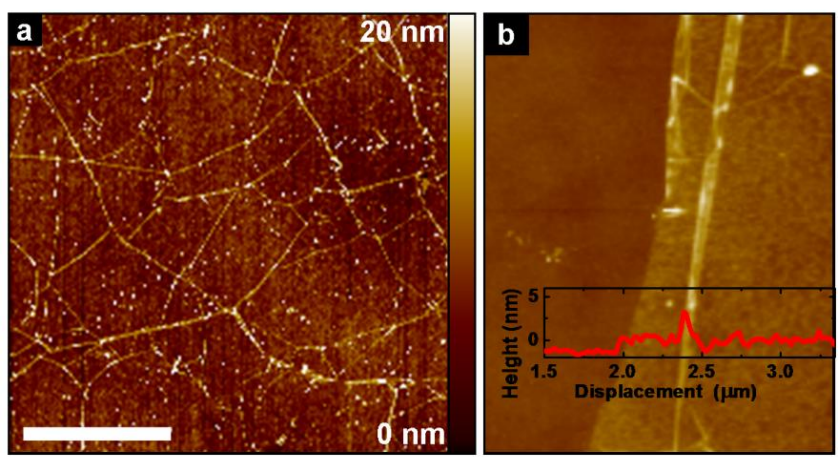

**Figure S1.** (**a**) Large scale AFM image of graphene coated Si/SiO$_2$. The graphene surface shows small ripples and few protrusions but is uniformly only one atom thick. Scale bar is 10μm. (**b**) AFM image of a partially graphene covered Si/SiO$_2$ substrate. A step like increase in thickness of ~1nm is due to the presence of graphene.

The surface roughness and Young's modulus of the various substrates used for these studies are summarized in Table S1.

| Substrate | $R_a$ (nm) | $R_q$ (nm) | $R_z$ (nm) | $E$ (MPa) (typical literature values) |
|---|---|---|---|---|
| Si/SiO$_2$ | 0.21 | 0.27 | 4.27 | 70x10$^3$ |
| PET | 0.47 | 0.66 | 7.66 | 1-15x10$^3$ |
| PDMS | 0.49 | 0.66 | 6.52 | 0.2-1 |
| PDMS with graphene | 0.63 | 0.87 | 14.3 | 3 |
| CVD Graphene on Si/SiO$_2$ (Bi-layer) | 0.41 | 0.78 | 7.56 | 1x10$^6$ |
| Carbon coated glass slide | 0.79 | 1.28 | 13.4 | 70x10$^3$ |
| Glass slide | 1.01 | 1.46 | 11.9 | 70x10$^3$ |

**Table S1.** Substrates overview. Roughness was measured by AFM and evaluated by three parameters: Average deviation from mean ($R_a$), Root Mean Square deviation ($R_q$) and the peak-to-peak distance ($R_z$). Young's Modulus E corresponds to typical literature values.

# Cell viability study

Four graphene-coated Si/SiO$_2$ chips, four uncoated Si/SiO$_2$ chips and eight cover slips (four as negative control and four as positive control) were put into 24 well plates. Trypan blue assay was used to count the number of viable hMSCs. By using the information obtained from the trypan blue assay, hMSCs were seeded at a density of 20,000 cells/ml into the above mentioned wells and incubated for 3 days in a humidified jacketed incubator (Binder), with a temperature of 37°C, 5% CO$_2$. On 2nd day 50µl of 1% sodium dodecyl sulphate (SDS) was added to positive control wells to induce cell death and continued with incubation.

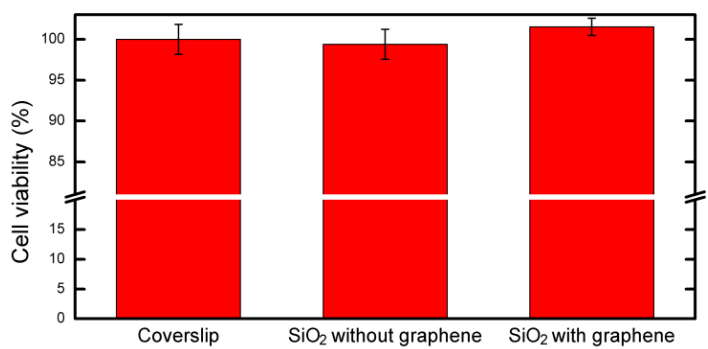

**Figure S2.** Graph showing the percentage of cell viability of hMSCs after 24 hours exposure to graphene-coated SiO$_2$ chips (right) in comparison to hMSCs grown on cover slips (left) and uncoated SiO$_2$ chips (center) as determined by MTT assay.

Assessment of cell viability was also determined using metabolic activity assays, 3-(4, 5-dimethylthiazol-2-yl)-2, 5-diphenyltetrazolium bromide (MTT) (Duchefa Biochemie). In this case, the test principle is based on the reduction of the tetrazolium salt from yellow color to purple color by the mitochondria present in viable cells (Mosmann, *J. Immun. Met.* **1983**, *65*, 55). After 3 days the 24 well plate was taken out from the incubator and all samples having cells growing on them were transferred to wells of another 24 well plate. 500 µl of tetrazolium salt (0.1 mg/ml) were added to each well including wells having no cells as blank and allowed to incubate for 4 hours in the jacketed incubator. After 4 hours, the contents in each well were pipetted out and 500 µl of DMSO were added to lyse the cells and to solubilize the insoluble formazan dye. The plate was shaken for 3 minutes to ensure homogeneity of color and aid solubilization of the dye prior to measure absorbance. The absorbance was recorded using the Benchmark plus Micro plate Spectrophotometer (Bio-Rad Laboratories) at 590 nm. The absorbance recorded is directly proportional to the number of viable cells present. All the experiments were repeated in triplicates and the results were expressed as the mean value. The cell viability was calculated in the following way:

$$\text{Cell viability} = \frac{\text{Absorbance of test - Absorbance of blank (PBS)}}{\text{Absorbance of control - Absorbance of blank (PBS)}}$$

Student t-test was used to calculate p-value. If the p-value was found to be less than the threshold (0.05) chosen for statistical significance, then the null hypothesis (which states that the two groups do not differ) was rejected in favor of an alternative hypothesis, which states that the groups do differ significantly. As seen from Fig. S2, there was no significant difference ($p>0.05$) in the percentage of cell viability of hMSCs growing on graphene-coated $SiO_2$ chips in comparison to hMSCs growing on uncoated $SiO_2$ substrates and cover slips. This result suggests that graphene is non-cytotoxic. The percentage of cell viability data as found from the above MTT assay was comparable to the percentage of cell viability as determined by cell viability by image analysis (as shown in the main text Fig. 2a).

## Experiment on flow cytometry

The evaluation by fluorescence-activated cell sorting (FACS) further supports the reported data by providing a quantitative evaluation of cell populations presenting specific characteristics. This represents, in our case, the ability of hMSCs to differentiate into osteogenic lineage once adequately stimulated. The hMSCs grown on different substrates (i.e. cover slips, uncoated-$Si/SiO_2$ and graphene-coated $Si/SiO_2$) were subjected to differentiation with osteogenic medium (in the presence or absence of BMP-2) and analyzed after 14 days by FACS. The harvested cells were fixed with 4% paraformaldehyde by incubating for 20 minutes. After centrifugation at 1500 RPM for 5 minutes and washing with PBS, the cell pellets were suspended in 100 mM glycine for 10 minutes to quench. The cells were then again centrifuged and washed with PBS and permeabilized by incubating in 50 µl of 0.1% Triton X for 30 minutes. Subsequently, the cells were washed with PBS and were incubated with mouse antihuman osteocalcin antibody for 30 minutes at room temperature. The cells were further washed with PBS and incubated with FITC conjugated goat anti mouse IgG for another 30 minutes. Finally the cells were washed 2-3 times with PBS and were analyzed using BD LSR II flow cytometer (Becton Dickinson).

The obtained curves are summarized in Fig. S3 and confirm the results obtained with Alizarin Red quantification (see main text Fig. 4). The FACS histogram reported below displays a single measurement parameter (relative fluorescence intensity due to FITC) on the x-axis and the number of events (cell count) on the y-axis. While the osteocalcin positive cell population representing osteogenically differentiated hMSCs can be found in form of single distinct shifts beyond relative fluorescence intensity of $10^3$, the undifferentiated hMSCs can be found in form of single histogram below such value.

As expected, negligible osteocalcin positive cells were found in case of hMSCs on substrates incubated in normal medium (histogram a). The expression of osteocalcin was maximal for all the substrates in osteogenic media with both graphene and BMP-2. This is similar to the results obtained with the alizarin red quantification and confirms the synergistic effect when both graphene and BMP-2 were concurrently present. Interestingly, osteogenic medium with graphene, but in the absence of BMP-2, reached almost the same levels of cell differentiation (83%) as those in osteogenic medium with both graphene and BMP-2 (100%). Taken together, both Alizarin Red Quantification and FACS analysis provide a quantitative proof on the ability of graphene to accelerate stem cell differentiation even in the absence of additional growth factor.

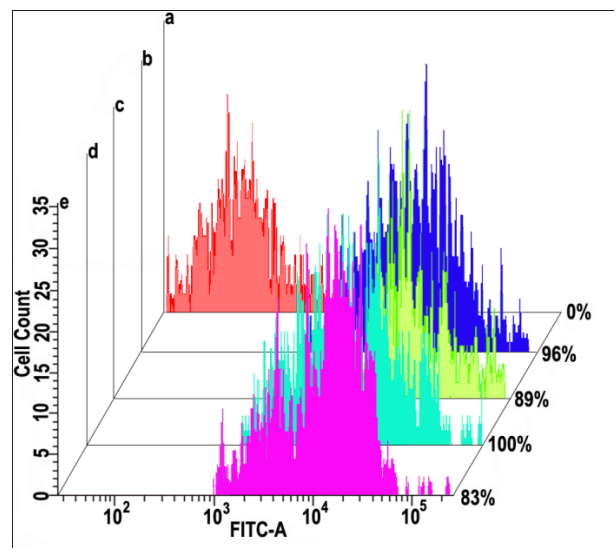

**Figure S3:** FACS analysis after 14 days of incubation of hMSCs. Curve **a (red)**: Cover slips without osteogenic media (OM) and without BMP-2 (negative control, 0% of differentiation); Curve **b (dark blue)**: Cover slips with both OM and BMP-2 (96% of cell differentiation); Curve **c (green)**: Si/SiO$_2$ with both OM and BMP-2 (89% of cell differentiation); Curve **d (light blue)**: Graphene with both OM and BMP-2 (synergistic effect, 100% of cell differentiation); Curve **e (magenta)**: Graphene with only OM, without BMP-2 (83% of cell differentiation).

# Control experiments with amorphous carbon

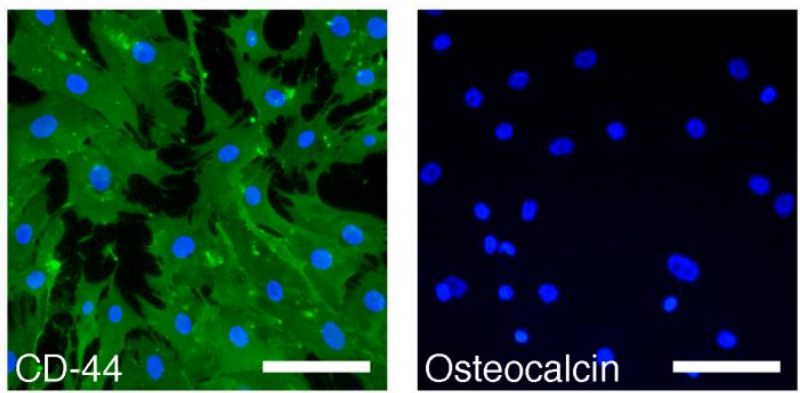

**Figure S4.** Immunostaining of cells growing on amorphous carbon-coated substrates. CD-44 (**left**) was still fully visible, while osteocalcin (**right**) was not visible. Scale bars are 100μm.

The osteogenic effect could in principle also take place in the presence of continuous thin carbon films. To verify the significance of graphene, substrates were coated with a thin layer of carbon using a JEE-420 vacuum evaporator. The thickness of the carbon coating was determined by AFM to be approximately 10 nm. These samples were then subjected to immunostainng of CD-44 and osteocalcin, similar to previous experiments. The carbon coated samples did not show a decrease in CD-44-fluorescence (Fig. S4 left), indicating no sign of differentiation. Equally important, no osteocalcin was stained (Fig. S4 right), confirming that no cell differentiation towards osteoblasts occured. Thus, a thin layer of carbon is insufficient to render stem cells osteogenic.

# Additional material parameters potentially responsible for accelerated stem cell differentiation in graphene coated substrates.

### a) Influence of surface chemistry

The role of defects and surface functionalization in CNTs has been heavily studied, but still remains not fully elucidated. The situation may in fact be more complex and may in part be also due to an increase in wettability of the CNT surfaces after plasma treatment. An increase of hydrophylicity is well correlated with an improved adhesion of hMSCs onto the substrates (Baik *et al.*, *Small,* **2011**, *7*, 741; Martins *et al.*, *Small*, **2009**, *5*, 1195).

Pristine graphene is known to be hydrophobic, however we see no degradation in adhesion compared to samples without graphene and observe an increase in differentiation instead. On the other hand, the chemicals and solvents involved in the preparation of graphene coated substrates could create either functionalized groups physisorbed on the graphene surface or change the hydrophobicity. Furthermore, water molecules and hydroxyl radicals trapped between graphene and substrate during the coating process can also affect the surface energy. The situation becomes even more complex, once we introduce the various media for cell culturing.

### b) Influence of surface polarization

Bodhak *et al.* (*Acta Biomaterialia*, **2009**, *5*, 2178) observed that polarized hydroxyapatite substrates have lower contact angle and better wettability. Negatively polarized substrates immersed in simulated body fluid showed enhanced bone like apatite formation compared to unpoled substrates, while positively poled substrates determined direct inhibition of apatite layer. Similarly, human fetal osteoblasts reported good adhesion and proliferation on negatively poled substrates where the positively poled substrate inhibited cell growth. Graphene is generally observed to be positively charged due to environmental doping. However, the doping level of graphene is on the level of +0.2μC/cm$^2$ and hence, much smaller than the doping levels at which a negative result was observed (+4.3 μC/cm$^2$). This may explain why we do not see any issues with cell growth or adhesion on graphene. This hypothesis could be checked in future experiments by, e.g. depositing graphene on ferroelectric PVDF films of different polarization corresponding to a range of charge density ranging from -3 μC/cm$^2$ to +3 μC/cm$^2$ .

### c) Influence of thermal conductivity

Thermal properties of graphene could in principle play a role in stem cell differentiation. However, the absence of differentiation on (HOPG) graphite demonstrates that this property can only provide a minor contribution. While graphene does have the highest thermal conductivity among the carbon allotropes, what really matters is the thermal conductance, which is orders of magnitude higher in graphite. The thermal conductance of bulk graphite is much higher than what the surface can provide.

### d) Influence of electrical conductivity

It has been shown that the electrical conductivity of CNTs does promote nerve cells' growth (Hu *et al. Nano Lett.*, **2004**, *4*, 507; Hu *et al. J. Phys. Chem. B*, **2005**, *109*, 4285; Malarkey *et al. Nano Lett.* **2009**, *9*, 264). Based on these results, it has been speculated that higher electrical conductivity could also lead to better differentiation (Voge *et al. J. Neural Eng.* **2011**, *8*, 011001; Chao *et al. Biochem. Biophys. Res. Commun.*, **2009**, *384*, 426). However, to the best of our knowledge so far no experimental evidence has been provided showing that electrical conductivity itself can lead to accelerated stem cell differentiation. This would also be consistent with our observation that HOPG graphite does not lead to bone cell differentiation. Here it is important to note, that while the electrical conductivity of graphene exceeds the electrical conductivity of graphite, what really matters is the electrical conductance. Similar to the thermal conductance, also here the electrical conductance of bulk graphite is higher than that of graphene.

## Graphene quality after cell removal

It is important to know whether the graphene sheet remains intact during the cell differentiation process over 14 days. In Fig. S5 we show a series of Raman spectra and corresponding optical images addressing this question. Fig. S5b shows the Raman spectra of graphene on Si/SiO$_2$ after cell removal. The residual material provides a strong background signal. Super imposed on this are small but discernable Raman peaks. As a reference, we show a sample without graphene, which underwent the same process. The corresponding small Raman specific peaks are clearly absent. We can almost fully remove the background after leaving the samples (shown in Fig. S5a) for 24 h in Acetone. The Raman spectra of the resulting sample clearly show the G and 2D peaks, which represent the Raman "fingerprints of graphene". Note also, that the absence of the D-peak at 1350cm$^{-1}$ indicates the lack of defects to the graphene crystal lattice (Ferrari *et al. PRL*, **2006**, *97*, 187401). The optical images also clearly shows that the graphene sheet remains largely intact.

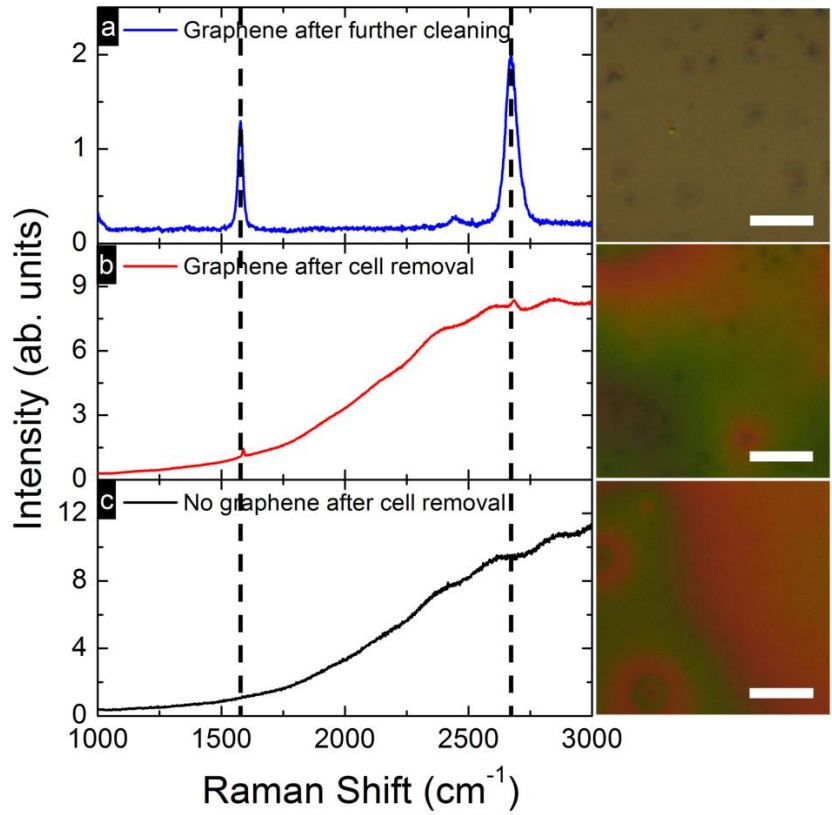

**Fig. S5**: Raman spectrum and optical images of (**a**) graphene on Si/SiO$_2$ after removal of cells and cleaning with acetone, (**b**) graphene on Si/SiO$_2$ after removal of cells and (**c**) Si/SiO$_2$ after removal of cells. Scale bars are 10µm.